\documentclass[12pt,a4paper,hyper]{article}
\pdfoutput=1
\usepackage{jheppub}
\usepackage{slashbox}
\usepackage{epsfig}
\usepackage{multirow}
\usepackage{url}
\usepackage{bbm}
\usepackage{graphicx}
\usepackage[utf8]{inputenc}
\usepackage[T1]{fontenc}
\usepackage{amssymb,amsfonts,amsmath,mathtools}

\font\cmss=cmss10
\font\cmsss=cmss10 at 7pt
\font\manual=manfnt

\newcommand{\bi}{\begin{itemize}}
\newcommand{\ei}{\end{itemize}}

\newcommand{\bea}{\begin{eqnarray}}
\newcommand{\eea}{\end{eqnarray}}
\newcommand{\be}{\begin{equation}}
\newcommand{\ee}{\end{equation}}
\newcommand{\ben}{\begin{eqnarray*}}
\newcommand{\een}{\end{eqnarray*}}
\newcommand{\bem}{\begin{pmatrix}}
\newcommand{\eem}{\end{pmatrix}}
\newcommand{\bl}{\begin{align}}
\newcommand{\el}{\end{align}}
\newcommand{\beg}{\begin{gather}}
\newcommand{\eeg}{\end{gather}}

\newcommand{\cA}{\mathcal{A}}

\newcommand{\cH}{\mathcal{H}}

\newcommand{\cN}{\mathcal{N}}

\newcommand{\cS}{\mathcal{S}}

\newcommand{\cZ}{\mathcal{Z}}

\newcommand{\bC}{\ensuremath{\mathbb{C}}}

\newcommand{\bK}{\ensuremath{\mathbb{K}}}

\newcommand{\bM}{\ensuremath{\mathbb{M}}}

\newcommand{\bR}{\ensuremath{\mathbb{R}}}

\newcommand{\bZ}{\ensuremath{\mathbb{Z}}}

\newcommand{\apm}{\alpha'}

\newcommand{\tS}{{\tilde S}}

\newcommand{\IH}{\mathbb{H}}

\newcommand{\IT}{\mathbb{T}}
\newcommand{\II}{\mathbb{I}} 

\renewcommand{\b}{\beta}

\renewcommand{\l}{\lambda}
\newcommand{\m}{\mu}

\newcommand{\s}{\sigma}                                   
\renewcommand{\t}{\tau}

\newcommand{\D}{\Delta}

\renewcommand{\O}{\Omega}

\newcommand{\vt}{\vartheta}

\newcommand{\TrH[1]}{ {\raise -.5em
                      \hbox{$\buildrel {\textstyle  {\rm Tr } }\over
{\scriptscriptstyle \cH _ {#1}}$}~}}
\newcommand{\res[1]}{{\raise -.5em 
\hbox{$\buildrel{\textstyle{\rm Res}}\over {\scriptscriptstyle {#1}}$}}}
\newcommand{\tends[1]}{{\raise -.5em 
\hbox{$\buildrel{\longrightarrow}\over {\scriptscriptstyle {#1}}$}}}
\newcommand\mygenfrac[2]{\genfrac{}{}{0pt}{}{#1}{#2}}

\newcommand{\sN}{\scriptscriptstyle N}
\newcommand{\half}{\frac{1}{2}}

\newcommand{\Tr}{\mbox{Tr}}

\def\dbend{\lower3.5pt\hbox{\manual\char127}}

\def\IL{\relax{\rm I\kern-.18em L}}
\def\IH{\relax{\rm I\kern-.18em H}}
\def\rlx{\relax\leavevmode}

\def\ZZ{\rlx\leavevmode\ifmmode\mathchoice{\hbox{\cmss Z\kern-.4em Z}}
 {\hbox{\cmss Z\kern-.4em Z}}{\lower.9pt\hbox{\cmsss Z\kern-.36em Z}}
 {\lower1.2pt\hbox{\cmsss Z\kern-.36em Z}}\else{\cmss Z\kern-.4em
 Z}\fi}

\title{\center {Quantum Entanglement in String Theory}}

\preprint{}
\author{
 Atish  Dabholkar\\
\leftline{}
\it {International Centre for Theoretical Physics\\
Strada Costiera 11, Trieste 34151 Italy}\\
}

\abstract{We define entanglement entropy in string perturbation theory using the orbifold method -- a stringy analog of the replica method in field theory. To this end, we use the Newton series to analytically continue in $N$ the partition functions for string orbifolds on $\mathbb{C}/\mathbb{Z}_N$ conical spaces,  known for all odd integer $N$. In the concrete example of ten-dimensional Type-IIB strings,  the one-loop partition function can be computed explicitly  and the  one-loop entropy can be expressed as a manifestly modular invariant  series in terms of the Weierstrass $\wp$ function. The convergence of the series is not evident but, from physical arguments based on holography, it is expected to yield a finite answer together with the tree level contribution.  This method has a natural generalization to other string compactifications and to higher genus Riemann surfaces; it can provide a modular invariant definition of generalized entropy in a given string vacuum to all orders, of potential interest for the generalized second law of thermodynamics. 
\vspace{5mm}
}


\begin{document}

\maketitle

\section{Introduction}

Our aim is to give a  definition of entanglement  entropy in Type-II superstring theory in Minkowski spacetime using an analytic continuation of the partition functions of the string  orbifolds constructed in \cite{Dabholkar1995c} which can be related to  Euclidean Rindler spacetime. The entropy thus defined in Rindler spacetime  has two possible physical interpretations -- either as quantum entanglement entropy or as  a quantum correction to black hole entropy. The two are expected to be related since the near horizon geometry of the two-sided eternal Schwarzschild black hole is approximated by Rindler spacetime.
If a proper  definition of Rindler entropy exists within string theory, then it would offer a way to talk about entanglement entropy in a consistent, ultra-violet finite theory of quantum gravity. 
Alternatively, it can teach us something about the quantum corrections to the Bekenstein-Hawking entropy  for generic nonsupersymmetric black holes in a generic string theory background. 
Both these interpretations are very interesting. 

In the context of $AdS/CFT$  holography, the  entanglement entropy in the boundary conformal field theory to leading order is related   by the Ryu-Takayanagi entropy formula to the area of the minimal surface in the quantum gravity dual theory in the bulk. The quantum correction to the Ryu-Takayanagi formula \cite{Ryu2006, Hubeny2007}  in the bulk dual has been argued to be given  by the quantum contribution to the bulk entanglement entropy \cite{,Lewkowycz2013a,Barrella2013,Faulkner2013,Engelhardt2015,Jafferis2016}. It is thus desirable to compute the entanglement entropy in string theory also from this perspective to better understand the relation between entanglement in the bulk and the boundary.  

Entanglement entropy and related notions such as relative entropy have come to play a central role in quantum information theory  in quantum mechanics and quantum field theory \cite{Nielsen2002, Preskill1998, Witten2018b,Calabrese2004,Casini2007,Calabrese2009} and more recently in the context of black hole physics and holograpy \cite{Rangamani2017} and in topics at the interface such as the proof of the Bekenstein bound \cite{Casini2008}. Finiteness of entanglement entropy in quantum gravity is at the heart of the information paradox and related issues in black hole physics. For example,   the formulation of the strong sub-additivity paradox for Hawking emission \cite{Mathur2005, Mathur2009, Almheiri2013} requires a notion of entanglement in a gravitational context. Similarly, the proof of the generalized second law of thermodynamics to one loop order \cite{Wall2012}
requires one to separate the relative entropy into the `entropy term' which captures the entanglement of the modes outside the black hole and the `energy term' which enters the Raychaudhuri equation. Each of these terms is individually ultraviolet divergent in perturbative quantum field theory but is expected to be finite in a consistent quantum theory of gravity such as string theory.  Thus, it is clearly desirable to arrive at an appropriate notion of entanglement entropy within string theory that is modular invariant and finite at one-loop order and possibly to all orders in perturbation theory. 

With these motivations, we review strings in Rindler spacetime in \S\ref{sec:Rindler} and outline the orbifold method for computing the entanglement entropy. For concreteness, we consider ten-dimensional Type-II strings on a conical space  with opening angle $2\pi/N$ viewed as a $\mathbb{C}/\mathbb{Z}_N$ orbifold, and construct the one-loop partition function in \S\ref{sec:Cone}. After reviewing the Newton series in  \S\ref{sec:Newton}, we use it in \S\ref{sec:Quantum} to analytically continue in $N$ the partition functions of the orbifolds to  obtain a series expansion for the quantum entanglement entropy. In \S\ref{sec:Classical} we discuss the classical contribution to the entanglement entropy. At tree level,  the worldsheet partition function vanishes. However,  there is boundary contribution to the spacetime free energy which can be computed using the fact that the tree level dilaton equations of motion are satisfied exactly for the string orbifolds \cite{Dabholkar2002}. This allows one to compute both the  classical and the quantum contributions to entanglement entropy uniformly using the orbifold method. 

At one-loop order, the formula \eqref{q-entropy} for  quantum entanglement entropy can be evaluated explicitly for the ten-dimensional Type-IIB strings. It yields an intriguing modular invariant formula as a  series involving the Weierstrass $\wp$ function. It is not evident if this series can be summed in some way, but one expects that the one-loop entropy should be finite by the following physical reasoning in the context of $AdS/CFT$ holography. The near horizon geometry of a Schwarzschild black hole in anti de Sitter spacetime is well approximated by Rindler spacetime. In the two-sided boundary conformal field theory living on the left and the right boundary, the black hole  is represented \cite{Maldacena2001a,VanRaamsdonk2010,Maldacena2013} as the thermofield double state $|\overline\O\rangle$ in the product Hilbert space $\overline \cH_L \otimes \overline \cH_R$ where $\overline \cH_L$ and $\overline \cH_R$ are the  left and the right boundary Hilbert spaces respectively\footnote{We use bar to distinguish all boundary quantities from the bulk quantities.}. Tracing the pure density matrix for the ground state over the left Hilbert space gives the thermal density matrix $\overline \rho_R$ on the right Hilbert space:
\begin{equation}\label{right-density}
	\overline\rho_R	= {\raise -.5em
                      \hbox{$\buildrel {\textstyle  {\rm Tr } }\over
{\scriptscriptstyle \overline \cH_L }$}~}
	 |\overline\Omega\rangle \langle \overline \O |
\end{equation} The von Neumann entropy $\overline S$ of this density matrix  is finite at each order in a large $N$ expansion. Hence, one expects that the corresponding bulk quantity $S$  must also be finite order by order in string perturbation theory.  A detailed analysis of the finiteness of the entropy \eqref{entropy} will be presented elsewhere \cite{DabholkarWiP}. 

These physical considerations have a more sophisticated formulation using von Neumann algebras of local observables which are natural from the perspective of  the boundary conformal field theory \cite{Leutheusser2021,Witten2021b,Leutheusser2021a,Witten2021a}. The algebra of local observables in the right conformal field theory is a Type-I von Neumann algebra. This is related to the fact that  $\overline S$ is finite. By contrast,  the bulk entanglement entropy in the naive field theory limit of string theory is divergent corresponding to the fact that the von Neumann algebra of observables of a local quantum field theory is Type-III. Holography implies that in a consistent quantum string dual in the bulk, this situation must be ameliorated leading to finite entanglement entropy, even though there is no natural notion of algebra of local observables in quantum gravity. We discuss this perspective in \S\ref{sec:Finite}. 

\section{Strings on Rindler Spacetime \label{sec:Rindler}}

Consider string theory on a ten-dimensional spacetime $\bM_{10} = \Sigma_{8} \times \bM_{2}$ where $\Sigma_{8}$ is an arbitrary Euclidean eight-manifold and $ \bM_{2}$ is two-dimensional Minkowski spacetime with metric 
\be
ds^{2} = -dT^{2}  + dX^{2}  \, , 
\ee
in the usual Minkowski coordinates $T, X$ with $-\infty < T < \infty$ and  $-\infty < X < \infty$. 
A uniformly accelerated observer in $\bM_{2}$ uses instead the Rindler coordinates  $(t, r)$ related to the Minkowski coordinates by
\be
T = r \cosh t \, , \quad X = r \sinh t  \,  
\ee
with $ - \infty < t < \infty$ and $0 \leq r < \infty$,  which cover only the right diamond in Minkowski space. 
In these coordinates,  the metric takes the form
 \be
ds^{2} = - r^{2 }dt^{2}  + dr^{2} \, .
\ee
Euclidean Rindler space is obtained by the Wick rotation $t = -i\theta$. The Euclidean Rindler metric is  thus identical to the flat Euclidean metric in polar coordinates
\begin{equation}%
ds^2 = r^2 d\theta^2 + dr^2 \, \qquad 0\leq \theta \leq 2\pi \, . 
\label{polar}
\end{equation}%
The   metric  is smooth at the origin if the angular coordinate has periodicity $2\pi$. 

As usual, periodicity  of the Euclidean time $\theta$ is interpreted as the inverse temperature. This implies that in the Minkowski vacuum, an observer with unit proper acceleration sees a thermal bath of strings at inverse Rindler temperature $2\pi$. The \textit{spacetime} partition function   is given by
\be
\hat Z = \Tr \exp\left({{-2\pi H_{R} }}\right) \, .
\ee
where $H_{R}$ is the Rindler Hamiltonian which generates the translations of Rindler time $t$. 
The partition function  can be represented formally as 
 a string field theory path integral on $ \Sigma_{8} \times \bR_{2}$ where $\bR_{2}$ is  the flat two-dimensional Euclidean space. 

 The thermal entropy $S$ of the Rindler heat bath is given, as usual, by a derivative of the partition function. 
Thus, to compute the Rindler entropy we need to consider the partition function at  inverse 
temperature $\beta$ slightly away from $2\pi$ to 
\begin{eqnarray}
\b := 2\pi - {\delta} :=    \frac{2\pi}{N }\, .
\end{eqnarray}
where 
$N \geq 1$ is continuous parameter and $N=1$  corresponds to flat
space. The partition function
\be\label{partfun}
\hat Z(N) := \Tr \left[ \exp \left({-\frac{2\pi}{N} H_{R}}\right) \right]
\ee
 is then represented as a Euclidean path integral on $\Sigma_{8} \times \bK_{N}$ where $\bK_{N}$ is a conical space with Euclidean metric
\bea\label{conementric}
ds^2 = r^2 d\theta^2 + dr^2 \, \qquad 0\leq \theta \leq \frac{2\pi}{N} \, . 
\label{cone}
\eea
The Rindler entropy is then given by
\bea\label{rindentro}
S = -\b \frac{\partial (\log \hat Z)}{\partial \b} +\log \hat Z.
\eea
evaluated at the Rindler temperature, or equivalently by
\bea \label{entropy}
S = \frac{d (N \, \log {\hat Z(N)} ) }{dN} \bigg|_{N=1} .
\eea
  
The range of the angular  coordinate  in \eqref{conementric} implies  a  deficit
angle $\delta_{N}$
\bea\label{deficit}
 \qquad \delta_{N} = 2\pi (1-\frac{1}{N}) \, 
\eea
and a  conical curvature singularity at the origin.  The Ricci scalar is given by a delta function
\bea
R(x) = 4\pi (1-\frac{1}{N}) \,  \delta^{(2)} (x) \, .
\eea
A conical space with a curvature singularity is not Ricci flat and  Einstein equations are not satisfied  unless there is a delta-function energy source at the tip of the cone. 
It would thus seem that string equations of motion would also not be satisfied.   If this were the case, then  it would be necessary to use an  off-shell formulation of string theory to compute the partition function and it would be a considerably difficult problem  to make sense of Rindler entropy in string theory. 

One can instead take an indirect route  \cite{Dabholkar1995c} using the fact that string propagation on conical spaces $\bK_{N}$ with deficit angle given by \eqref{deficit} for arbitrary \textit{odd integer} $N \geq 1$   is well defined. In this case, one can regard   $\bK_{N}$ as a $\bZ_{N}$ orbifold \cite{Dixon1985} of the Euclidean space $\bR_{2}$ or equivalently of  complex plane $\bC$. String theory on these  $\bZ_{N}$ orbifolds can be constructed using worldsheet orbifold techniques. 
Conformal invariance of the worldsheet sigma model implies, somewhat surprisingly, that  the classical string equations of motion are satisfied \cite{Dabholkar2002}. 

Given $\hat Z(N)$ for all integer values of $N\geq 1$, one can try to  find an analytic continuation valid in the complex $N$ plane.  We will assume certain analyticity conditions required by Carlson's theorem \cite{Boas1954} so  that the resulting analytic continuation is unique. Given an analytic expression for  $\hat Z(N)$, using \eqref{entropy} one can then find the Rindler entropy.  This is the approach we will follow. For earlier work see \cite{He2015, Wittenb}. 

The orbifold method of analytically continuing from  deficit opening angles $2\pi /N$ is a variant of the replica method \cite{Mezard1986} of analytically continuing from  surplus opening angles $2\pi \cN$ for an integer $\cN \geq 1$.  See \cite{Wittenb} for a detailed discussion of the relation betweeen the replica method and the orbifold method with $\cN = 1/N$. In string theory, an obvious  advantage of the orbifold method as mentioned above is that the orbifolds are saddle points of the string field theory path integral and   the  orbifold partition functions can be constructed  relatively easily. 

It is convenient for later discussion to write  the  entropy as
\begin{equation}
	S = S^{(0)}	+ S_q
\end{equation}
where $S^{(0)}$ is the classical tree-level contribution  and $S_q$ is the quantum loop  contribution from higher  genus  Riemann surfaces with $g\geq 1$. 

At the quantum level, the spacetime partition function $\hat Z_q(N)$ of the $\bZ_{N}$ orbifold  theory is related to the worldsheet partition function   $Z_q(N)$  by the relation
\be\label{spaceworld}
\log (\hat Z_q(N))  = Z_q(N) \, . 
\ee
with a genus expansion
\begin{equation}
	Z_q(N) = \sum_{g=1}^\infty Z^{(g)}(N) \, .
\end{equation}
The quantum entanglement entropy then has the corresponding expansion
\be
S_q = \frac{\partial}{\partial N} \left( N Z_q(N) \right) \bigg|_{N=1} \, = \sum_{g=1}^\infty S^{(g)}. 
\ee

At the classical level, the relation  \eqref{spaceworld} does not hold. The worldsheet partition function vanishes on the conical backgrounds. On the other hand, one can argue that the logarithm of the the classical partition function $\hat Z^{(0)}(N)$ does not vanish and receives a  nonzero boundary contribution. One can show \cite{Dabholkar2002} by analytically continuing $\log (\hat Z^{(0)}(N))$ in $N$   that $S^{(0)}$ is simply the Bekenstein-Hawking entropy:
\begin{equation}
	S^{(0)} = \frac{A}{4G}
\end{equation}
Thus, the total entanglement entropy in string theory is given by 
\begin{equation}\label{S-total}
	{S}  = \sum_{g=0}^\infty S^{(g)} \, = \frac{A}{4G} + S_q \, .
\end{equation}

As mentioned earlier, the quantum entanglement entropy $S_q$ admits two physical interpretations which \textit{a priori} are distinct. 

\subsection*{\hspace{2pt} $\bullet$ \hspace{2pt} \textbf{ Quantum Entanglement Entropy} }

One can imagine dividing space at $X=0$ into left and right sides separated by 
the surface $\Sigma_{8}$. 
The Rindler Hamilotonian  can be interpreted in the Tomita-Takesaki theory \cite{Takesaki1970,Borchers2000} as  the modular Hamiltonian corresponding to the `cyclic and separating' Minkowski vacuum state for the algebra of local observables restricted to the right side. See \cite{Witten2018} for a review and relevant references.  The unnormalized density  matrix $\rho$ obtained by tracing over the left wedge is then given formally by the thermal density matrix for the Rindler Hamiltonian: 
\be \label{Rindlerrho}
\rho = \exp\left({{-2\pi H_{R} }}\right) \, .
\ee
The partition function \eqref{partfun} can therefore be viewed as  the Renyi entropy 
\be
\Tr (\rho^{\cN} )
\ee
with $\cN = 1/N$
if it can be appropriately analytically continued to integer $\cN$. The Rindler entropy can thus be identified with the von Neumann entropy of the thermal density matrix or equivalently with  the entanglement entropy associated with the surface $\Sigma_{8}$. 

In quantum gravity one cannot define an algebra of local observables and it is not clear how to define notions like modular Hamiltonian and  entanglement entropy. However, in perturbation theory around a fixed background metric, one might hope to be able to give meaning to these concepts in the same way as in a local quantum field theory.
Moreover, given the ultra-violet finiteness of the theory, Rindler entropy  is expected to be  better defined in string theory. 

It has been argued that the field-theoretic divergences in the entanglement entropy \cite{Susskind1994,Kabat1995} can be absorbed into the Einstein-Hilbert action as the renormalization of Newton's constant. The quantum effective action for massless fields cannot  be summarized entirely by  local Wilsonian terms and includes highly nonlocal terms in addition to the Einstein-Hilbert term. One thus expects nonzero but finite contribution to the entropy even after the renormalization of Newton's constant has been taken into account. In other words, there is no reason in general why finite piece in the renormalization of Newton's constant will equal the finite piece in the entanglement entropy. Holographic considerations in \S\ref{sec:Finite} also lead one to expect finite but nonzero contributions to the entanglement entropy order by order in string perturbation theory.
 
\subsection*{\hspace{2pt} $\bullet$ \hspace{2pt}  Quantum Black Hole Entropy}

\vskip 10pt

In black hole spacetimes with bifurcate horizons, the near horizon geometry of the spacetime transverse to the horizon is a Rindler spacetime. Rindler temperature  is red-shifted to the inverse Hawking temperature
$\b_H={8\pi G M}$ for the asymptotic observer. Rindler entropy can then be interpreted as the  quantum correction to the black hole entropy in the limit of infinite horizon area \cite{Sorkin2014, Bombelli1986,Srednicki1993,Susskind1994,Callan1994a,Solodukhin1995,Fursaev1996,Solodukhin2011,Frolov1996}.  To be concrete,  one can imagine computing the entropy of a four-dimensional finite area black hole by performing the string field theory path integral in the black hole background. In this case,  one would expect an expansion for the entropy of the form
\be\label{expansion}
S = \frac{A}{ l_s^{2 }g^{2}}( \frac{1}{4} + a_{1} g^{2}  + \ldots) + \log (A/l_s^{2}) ( b_{1} g^{2}  + \ldots )  
+  \frac{l_s^{2}}{A} ( c_{0} + c_{1} g^{2}  + \ldots ) + \ldots
\ee
where $l_s$ is the string length and $g^{2}$ is the four-dimensional string coupling. The term logarithmic in $A$ arises from the loops of massless particles \cite{Sen2012} and the   terms suppressed by inverse powers of $A$ correspond to contributions to the Wald entropy from  higher derivative terms. In the Wilsonian effective action, such terms would  arise from integrating out massive string states including the string-loop corrections. In the large $A$ limit, the term proportional to $A$ will dominate.  Our one-loop computation could then be interpreted as a computation of the coefficient $a_{1}$.  In field theory, $a_{1}$ is ultraviolet divergent, but string theory is expected to provide a natural UV cutoff to yield a finite answer. 
\vspace{10pt}

\section{Type II Strings on a Cone \label{sec:Cone}}

As a  concrete example of the considerations in the previous section, we
consider Type-IIB string
 on $\bM_8 \times \bK_{N}$ where $\bM_{8}$ is $7+1$ dimensional Minkowski spacetime. 
We start with the Green-Schwarz string on  $\bM_8 \times \bC$ and  fix the light-cone gauge  by using
two of the  directions in $\bM_8$ to obtain a sigma model in flat space $\bR_6 \times \bC$. The Green-Schwarz Lagrangian is given by
\be
{\cal L} = - \frac{1}{\pi} \partial_{+} X^i  \partial_{-} X^i
 -\frac{i }{\pi}S^{a} \partial_{+} S^a
 -\frac{i }{\pi} \tS^{\dot a} \partial_{-} \tS^{\dot a} \ ,
\ee
where the coordinates $X^i$  transform in the  vector 
(${\bf 8}_v$) representation of $SO(8)$ and $S^a $ and $\tS^{ \dot a}$ transform respectively in  
the    spinor (${\bf 8}_s$)  and conjugate spinor   (${\bf 8}_c$ )
representations   of $Spin(8)$. 

The orbifold group $\bZ_N$ is a subgroup of  planar rotations of the complex plane $\bC$,  so
we  use the decomposition $Spin(8) \supset Spin(6) \times Spin(2)$, or
equivalently  $Spin(8) \supset SU(4) \times U(1)$. The Euclidean Rindler Hamiltonian is simply the generator $J$ of the $U(1)$ symmetry.
The vector and the spinor representations then decompose
as follows:
\bea\label{charges}
{\bf 8}_v &\rightarrow& {\bf 6} (0) + {\bf 1} (1)+ {\bf 1} (-1)\\
{\bf 8}_s &\rightarrow &{\bf 4} (\half) + {\bf {\overline 4}} (-\half)\\
{\bf 8}_c &\rightarrow& {\bf 4} (- \half) + {\bf {\overline 4}} (\half)
\eea
Here the numbers in the parentheses are the  $U(1)$ charges. 

The  fields charged under the $U(1)$ consist of one complex  boson $X$ with charge $1$,
eight  fermions $ S^m , \tS^m$ with charge
$\half$,  and their complex conjugates. The index $m$  transforms
in the  ${\bf 4}$ of $ SU(4)$. These are the only fields that get twisted by the  orbifold group. 
With the above conventions, the orbifold group $\bZ_{N}= \{1, g, \ldots, g^{N-1}\}$ is generated by\footnote{The $N$-th power of the naive choice $g := \exp {\frac{2\pi i J}{N}}$  gives $(-1)^F$, where $F$ is the spacetime fermion number. We want $g^N=1$ even while acting on the fermions. Otherwise, one would obtain a $\bZ_{2N}$ orbifold which can also be viewed as a $\bZ_{N}$ orbifold of a $\bZ_{2}$ orbifold.  The $\bZ_{2}$ orbifold has ten-dimensional tachyons and corresponds to  a different vacuum of string theory denoted as Type-0 theory.  We could equivalently consider $g := \exp {\frac{2\pi i J}{N}} (-1)^F$ as the generator, but it only relabels the sectors since $N$ is odd.}

\be\label{twist}
g := \exp {\frac{4\pi i J}{N}}
\ee

The orbifold partition function for these fields is of the form 
\be\label{orb-part}
\cZ  (\t , \overline \t ; N) =  \frac{1}{N} \sum_{k, l = 0} ^{N-1}\cZ_{k,l} (\t , \bar \t ; N) \, .
\ee
Each  term $\cZ_{k,l}$ is a partition function
for the fields $ X, S^m , \tS^m$ and their complex conjugates with
twisted boundary conditions
\bea
S^{m}(\s_1 +1 \, , \, \s_2)  &= 
e^{\frac{2\pi i k } {N} } \, S^{m}(\s_1 \, , \, \s_2)\, , 
\quad S^{m}(\s_1  \, , \, \s_2 +1)  &=  
e^{\frac{2\pi i l } {N} } \, S^{m}(\s_1 \, , \, \s_2)\, \nonumber \\
\tS^{m}(\s_1 +1 \, , \, \s_2)  &=  
e^{-\frac{2\pi i k } {N} } \, \tS^{m}(\s_1 \, , \, \s_2)\, , 
\quad \tS^{m}(\s_1  \, , \, \s_2+1)  &= 
e^{-\frac{2\pi i l } {N} } \, \tS^{m}(\s_1 \, , \, \s_2)\, , \nonumber \\
X(\s_1 +1 \, , \, \s_2)  &= 
e^{\frac{4\pi i k } {N} } \, X (\s_1 \, , \, \s_2)\, , 
\quad X(\s_1  \, , \, \s_2+1)  &= e
^{\frac{4\pi i l } {N} } \, X (\s_1 \, , \, \s_2)\, .
\eea
Comparing these boundary conditions in \cite{Alvarez-Gaume1986} we see that the fermions are twisted by
\be
a= \frac{k}{N} + \half \quad   and \quad  b= \frac{l}{N} + \half 
\ee
and the bosons are twisted by
\be
a= \frac{2k}{N} + \half \quad   and \quad  b= \frac{2l}{N} + \half  \, . 
\ee
Given these boundary conditions,  we can write down the one-loop partition function for the charged coordinates by inspection or by explicit oscillator computation \cite{Alvarez-Gaume1986}:
\bea \label{orb-part2}
\cZ (\t , \overline \t ;  N) =  \sum_{k, l = 0}^{N-1}
  \,  \Bigg|
\frac{  \vartheta ^4 \bigg[ 
{{ \mygenfrac{{\frac{k} {\sN} +\half}  }{{\frac{l} {\sN} +\half } }}\bigg] } (0|\t) }
{\eta^3 (\t) \vartheta \bigg[{ { \mygenfrac{\frac{2k} {\sN} +\half} {\frac{2l}
{\sN}+\half } } \bigg] } (0|\t) } 
\Bigg| ^{2} \, .
\eea

The total one-loop partition function including the six uncharged bosonic coordinates is
\be
Z^{(1)}(N) = \frac{A_{H}}{N} \int  \frac{d^{2}\t}{\t_{2}^{5}} \big|\frac{1}{\eta^{6}(\tau)}\big|^{2}\cZ (\t , \bar \t ; N)
\ee
where $A_{H}$ is the regularized horizon area measured in string units 
\begin{equation}
	A_H := \frac{V_8}{(2\pi l_{s})^8} \, ,\qquad 	 \qquad l_{s}^{2}= \apm. 
\end{equation} 
The superscript indicates that we are computing the one-loop partition function. 

The theta functions with characteristics can be related to the more familiar Jacobi theta function by the relation
 \be\label{relation0}
\vartheta \bigg[\mygenfrac{a +\half}{b+\half }\bigg] (\t) 
= e^{\pi i \, a^{2} \t + 2\pi i \, a\, (b + \half)}\,  \vartheta(a\t + b|\tau)
\ee
where the Jacobi theta function has the product representation
\begin{equation}\label{thetaproduct}
	\vt (z| \t) = -2 q^{\frac{1}{8}} \, \sin{\pi  z}
\prod_{n=1}^{\infty} (1-q^{n}) (1 - q^{n} y )\,
 (1 - q^{n} y^{-1}) \ .
\end{equation}

Using \eqref{relation0} we can write 
\begin{equation}
	\frac{1}{\eta^9(\tau)} \, \frac{  \vartheta ^4 \bigg[ 
{{ \mygenfrac{{\frac{k} {\sN} +\half}  }{{\frac{l} {\sN} +\half } }}\bigg] } (0|\t) }
{ \vartheta \bigg[{ { \mygenfrac{\frac{2k} {\sN} +\half} {\frac{2l}
{\sN}+\half } } \bigg] } (0|\t) } \, = \,   \frac{1}{\eta^9(\tau)} \, \frac{\vt^4(\frac{k\tau}{N} + \frac{l}{N}|\tau)}{\vt(\frac{2k\tau}{N} + \frac{2l}{N}|\tau)} 
\end{equation} 
It is convenient to define
\begin{equation}\label{Gdef}
	G(z|\tau) \, := \,
 \frac{1}{\eta^9(\tau)} \, \frac{\vt^4(z|\tau)}{\vt(2z|\tau)} 
 \end{equation}
The $G$-function so defined is a doubly  periodic and odd function on the torus:
\be
G(z|\t) = G(z+1|\t) = G(z+\t| \t) = - G(-z|\t) \,  
\ee
and transforms like a modular form
of weight $-3$:
\be\label{Hmod}
 G (-\frac{z}{\t}|- \frac{1}{\t}) = (- \t)^{-3} \, G(z|\t) \, \, .
\ee
The partition function can then be written compactly as
\be\label{one-loop}
Z^{(1)}(N) = \frac{A_{H}}{N} \int  \frac{d^{2}\t}{\t_{2}^{5}} 
\, \sum_{k, l = 0}^{N-1}
  \,  \bigg|G(\frac{k}{N} \tau + \frac{l}{N} \, | \t)\bigg|^2
\ee

The function $G(z|\t)$ is a meromorphic function over the elliptic curve or the genus-one Riemann surface $E(\tau)$ with the complex structure parameter $\t$. The elliptic curve can be thought of as the coset $\mathbb{C}/\mathbb{Z} \t + \mathbb{Z}$ of the complex plane $\mathbb{C}$ with coordinate $z$, modded out by the lattice $\mathbb{Z} \t + \mathbb{Z}$.  Since the  Jacobi theta function has a simple zero at all lattice points,  it is easy to see from \eqref{Gdef}  that $G(z|\t)$ has  triple zeroes at all lattice points $z = \l \tau + \m$ for $\l, \t \in \mathbb{Z}$ and  simple poles at the half lattice points.  
Knowing all its poles and zeroes, $G(z|\t)$ can be seen to be \cite{Wittenb} proportional to the inverse of the derivative of the Weierstrass $\wp$-function:
\begin{equation}\label{Gpe}
	G(z|\t) = -i  \frac{(2\pi)^3}{\wp’(z, \t)} \, .
	\end{equation}
Recall \cite{Chandrasekharan1985} that the  $\wp$ function is defined by the lattice sum
\begin{equation}
	\wp(z, \t) :=    \sum_{n, m \in \mathbb{Z}} \frac{1}{\left( z-m\t -n \right)^2 } \, -\sideset{}{'} \sum_{n, m \in \mathbb{Z}}\frac{1}{\left(m\t + n \right)^2 } \, .
\end{equation}
where the primed sum excludes the lattice point $n =m =0$. 
Its derivative 
\begin{equation}
	\wp'(z, \t) := \frac{d}{dz} \wp(z, \t) = - 2 \sum_{n, m \mathbb{Z}} \frac{1}{\left( z-m\t - n\right)^3}
\end{equation}
is an odd function in $z$. For fixed $\tau$, it has  triple poles at all lattice points and single zeros at all half lattice points.
Let $x = \wp(z, \t)$ and $y = \wp'(z, \t)$; then they satisfy the cubic equation 
\begin{equation}
	y^2 = 4 x^3 - g_4(\tau) x - g_6(\t) \, ,
\end{equation}
defining an elliptic curve with complex structure parameter $\tau$, where $g_4$ and $g_6$ are proportional to the Eisenstein series: 
\bea
g_4(\tau) = 60 \sum_{n, m \in \mathbb{Z}} \frac{1}{\left(m\t + n \right)^4 } \, , \qquad \qquad
g_6(\tau) = 140 \sum_{n, m \in \mathbb{Z}} \frac{1}{\left(m\t + n \right)^6 } \, .
\eea

\section{Newton Series\label{sec:Newton}}

Newton series is  an  analog in finite calculus of the  Taylor series in infinitesimal calculus. For any polynomial function and many analytic functions, it provides an exact series expansion. It is thus a natural interpolation to consider for our purpose of analytically continuing in $N$. 

Recall that for an analytic function $f(x)$, its Taylor series around $x=0$ is given by
\begin{equation}
	f(x) = \sum_{r=0}^\infty \frac{D^r[f](0)}{r!} x^r \, ,
\end{equation}
where 
\begin{equation}
	D^r[f](0) := \frac{d^r f}{dx^r}(0) \equiv f^{(r)}(0) \, 
\end{equation}
is the $r$-th derivative of the function evaluated at the origin. This expansion is valid in the domain of analyticity around the origin.   We are  interested in finding the function $f(x)$ for non-integer values of $x$ when we do not know these derivatives at the origin but  know only the values $\{ f(n) \}$ of the function  for all nonnegative integers $\{n\}$. One would like to construct an analytic function from this discrete data much as the Taylor series  constructs an analytic function from the discrete data $\{f^{(r)}(0)\}$.

To find such an interpolation, the  Newton series replaces the derivatives  by a natural definition of finite differences:
\begin{equation}\label{defdelta}
	\Delta^r[f](0) = \sum_{s=0}^r (-1)^{r-s} \frac{(r)_s}{s!} f(s)
\end{equation}
where $(r)_s$ is the $s$-th \textit{falling power} of r:
\begin{equation}
	(r)_s := r (r -1) (r-2) \ldots (r-s+1) = \frac{\Gamma(r+1)}{\Gamma(s +1)}
\end{equation}
The  Newton series  is then defined by
\begin{equation}\label{defNewton}
	f(x) := \sum_{r=o}^\infty \frac{\D^r[f](0)}{r!} \,  (x)_r \, .
\end{equation}
with  obvious formal analogy with the Taylor series. By construction, $f(x)$ thus defined takes values $f(n)$ at the non-negative integers $n$ but now interpolates for non-integer $n$.
 Convergence of the Newton series is in general not guaranteed.  If the Newton series converges, then Carlson theorem \cite{Boas1954} provides the necessary and sufficient conditions for the uniqueness of the resulting function $f(x)$. 
   
It is convenient to introduce an operator notation which makes  various aspects of the  Newton Series more transparent, especially the appearance of binomial coefficients in \eqref{defdelta}. 
Let $\IT_h$ be an operator of step size $h$ acting on functions as
\begin{equation}
	\IT_h[f](x) = f(x+h) \,. 
\end{equation}
Let $\II$ be the identity operator. 
Then the difference operator of step size $h$ can be defined by
\begin{equation}\label{defdifference}
	\Delta_h[f](x) = (\IT - \II)[f] (x) = f(x+h) -  f(x)
\end{equation}
In finite calculus, the difference operator above  is  a natural generalization  of the quantity  $ h D_h[f]$ in infinitesimal calculus where one takes the limit $h \rightarrow 0$ to define the differential operator $D[f]$. 
Powers of $\Delta_h$ can then be defined by the operator equation
\begin{equation}
	\Delta^r_h[f] = (\IT_h - \mathbb{I})^r[f]
\end{equation}
For our purposes, the  step size is unity, $h=1$, and in that case we drop the subscript. The definition \eqref{defdelta} then follows by  binomial expansion. 
It is also easier to see in this formalism that $f(x)$ defined by Newton series \eqref{defNewton} takes values $f(n)$ at $x=n$. If we substitute $x=n$ on the right hand side of \eqref{defNewton} then the infinite series terminates at $r=n$ and one obtains
\begin{equation}
	\sum_{r=0}^n \frac{\D^r[f](0)}{r!} \, (n)_r = \sum_{r=0}^n \frac{n!}{r!(n-r)!} \,  \D^r[f](0)
\end{equation}
which we recognize as the binomial expansion of $(\II + \Delta)^n[f]$. Hence the Newton series gives 
\begin{equation}
	f(x)\big|_n := (\II + \Delta)^n [f](0) = \IT^n[f](0) = f(n) \, 
\end{equation}
for all integer values $x=n$ as required. For more on calculus of finite differences see \cite{Milne-Thomson}.

This formulation also allows one to write the entropy formula in a more succinct form. The expression of the entropy takes the form
\begin{equation}\label{q-entropy1}
	S_q= \half \sum_{r=1}^\infty\frac{(-1)^{r+1}}{r} \Delta^r[f](0) \, .
\end{equation}
We recognize the infinite series as the Taylor expansion for $\log(\II + \Delta)$. Therefore, one can write
\begin{equation}
	S_q = \half \log (\IT) [f] (0)
\end{equation}
with $f(n) := Z(2n+1)$. 

To understand  the relation between the   differential operator $D$ and the difference operator $\D$ more generally, we Taylor expand \eqref{defdelta} in $h$ and set $h=1$,  to obtain
\begin{equation}
	\Delta [f] = \sum_{1}^\infty \frac{D^r}{r!} [f] = (e^D -\II) [f] \, , 
\end{equation}
or equivalently 
\begin{equation}
	D[f] = \log( \II +\Delta) = \log (\IT)[f] \, .
\end{equation}
As an operator equation it holds for polynomials and for functions for which the Newton series converges. The expression for the entropy above then follows. 
%

\section{Quantum Entanglement Entropy in String Theory \label{sec:Quantum}}

To obtain the entanglement entropy we apply the Newton interpolation formula to the function 
\begin{equation}
	f(n) := Z(2n+1) \qquad n=0, 1, 2, \ldots \, .
\end{equation}
to obtain the function for real $x$ near the origin. This enables us to express the derivatives at the origin in terms of the values $\{ f(n) \}$. Quantum entanglement entropy is then given by
\begin{eqnarray}
	S_q &=& \frac{d (N \, Z(N) }{dN} \bigg|_{N=1} = \frac{d \left((2n+1) \,  f(n)\right) }{d(2n+1)} \bigg|_{n=0} \\
	 &=&  f(0) + \half f’(0) \, 
\end{eqnarray}
where $f’(n)$ is the derivative with respect $x$ evaluated at $n$.
Assuming the convergence of the Newton series, the derivative of the interpolated function is given by
  \begin{equation}
	f’(x) := \sum_{r=o}^\infty \frac{\D^r[f](0)}{r!} \,  \frac{d(x)_r}{dx}\, .
\end{equation}

Since we are only interested in the derivative at the origin we note that
\begin{eqnarray}
	\frac{d(x)_r}{dx}\big|_{x=0} &=& \frac{d}{dx}\left[x (x-1)(x-2) \ldots (x + 1-r)\right] \\
	&=& (-1)(-2)(-3) \ldots	 (-(r-1)) = (-1)^{r-1}(r-1)! \, .
\end{eqnarray}
Using the fact that $f(0) = Z(1) =0$ we obtain
\begin{equation}\label{q-entropy1}
	S_q= \half \sum_{r=1}^\infty\frac{(-1)^{r-1}}{r} \Delta^r[f](0) \,.
\end{equation}
Substituting for $\Delta^r[f]$, we obtain
\begin{equation}\label{q-entropy}
	S_q = -\half \, \sum_{r=1}^\infty\frac{1}{r} \sum_{s=0}^r \, (-1)^s \, 
	\frac{(r)_s}{s!}\, Z(2s+1) \,.
\end{equation}

We have thus obtained an explicit expression for entanglement entropy as a  sum over the values of  partition functions of orbifolds. This formula is valid for any compactification and at any genus as long as the partition functions are known. Since the partition functions are manifestly modular invariant, the resulting entropy is also modular invariant. 

For concreteness, one can apply these general considerations to the orbifolds of ten-dimensional Type-IIB string constructed earlier. in this case, at one-loop order one can write an explicit expression for the entropy using  the partition function  \eqref{one-loop} as an integral over the moduli space:
\begin{equation} \label{entropy}
	S^{(1)} = A_{H}\int  \frac{d^{2}\t}{\t_{2}^{2}} \,  
	\cS^{(1)}(\tau)
	\end{equation}
where the entropy-density $\cS^{(1)}(\tau)$ is a nonholomorphic modular function over the moduli space:
\begin{equation}\label{entropy-density}
	\cS^{(1)}(\tau) = -\frac{(2\pi)^6}{2 \t_2^3} \, \sum_{r=1}^\infty\frac{1}{r} 	\, \sum_{s=0}^r \, (-1)^s \frac{(r)_s}{s!} \, \frac{1}{(2s+1)}\,\sum_{k, l = 0}^{2s}
  \, \bigg| \frac{1}{\wp’(\frac{k}{2s+1} \tau + \frac{l}{2s+1}, \t )} \bigg|^2 
 \,.
\end{equation}
where we have substituted  for $G(z|\t)$ in terms of the Weierstrass $\wp$ function using \eqref{Gpe}.

It is not known \emph{a priori} if this series is convergent, divergent, or asymptotic. It would  be interesting to examine if it can be summed explicitly, perhaps  with appropriate regularization, or  using other methods like Borel summation to extract the physical implications.
Since each of the partition functions has a tachyonic divergence in the integral over moduli space, it appears that the Rindler entropy would also inherit this divergence.  However,  one may hope that, first performing the sum and then performing the modular integral might ameliorate the divergence  as is indicated in the closed string channel in the analysis of open strings in Rindler spacetime \cite{Wittenb}.  As mentioned in the introduction, there are  physical reasons to expect that the entanglement entropy in string theory should be finite. We explain the reasoning in \S\ref{sec:Finite}. 

\section{Classical Entanglement Entropy in String Theory \label{sec:Classical}}

Its natural to ask if the  tree-level classical contribution  corresponding to the Bekenstein-Hawking entropy should be added by hand to obtain the total entanglement entropy \eqref{S-total} or if it can also be computed directly by the orbifold method.   This was investigated in \cite{Dabholkar2002} which we review below.  We will see that both the classical and quantum contributions to entanglement can be treated uniformly using the orbifold method by analytic continuation of \textit{spacetime} free energy for the orbifolds including appropriate boundary terms at tree level and using the fact that the orbifolds satisfy the classical string equations of motion exactly.  

If the naive relation \eqref{spaceworld} between the spacetime and worldsheet partition functions were valid even at tree level, then the tree-level entropy would  be zero.  However, as pointed out in \cite{Dabholkar2002}  there is a potential subtlety. For example,  the Schwarzschild metric also satisfies string equations of motion to leading order  and defines a conformally invariant sigma model. The bulk Einstein action for this solution is indeed zero but  the total action is nonzero because there is a boundary contribution coming from the Gibbons-Hawking boundary term \cite{Gibbons1977} . If the naive identification  \eqref{spaceworld}  were correct then one would  conclude that the tree level free energy and consequently the Gibbons-Hawking entropy of the Schwarzschild black hole is zero.  This is manifestly not correct. Thus the worldsheet partition function  could equal at most the \textit{bulk} spacetime action and need not capture the Gibbons-Hawking boundary contribution which one believes must exist from spacetime considerations.

Motivated by the Gibbons-Hawking computation, we  assume that the relation \eqref{spaceworld} is valid only for the bulk spacetime action. 
For the $\bZ_N$  orbifold conformal field theory, the equations of
motion for the dilaton and the graviton are satisfied exactly with a constant dilaton. The tree level spacetime action always comes multiplied by the factor $e^{-2\phi}$ where $\phi$ is the dilaton and thus acts as a source for the dilaton. This implies that 
 the  bulk action must vanish exactly because otherwise there would be a dilaton tadpole. The vanishing of the bulk spacetime action would be  consistent with the vanishing of the worldsheet partition function.  
 
 We would now like to deduce the boundary contribution to the tree level action following the Gibbons-Hawking procedure  around  these orbifold saddle points of the string field theory action. We do not know of a worldsheet computation to achieve this, so we will follow instead a spacetime approach as in \cite{Dabholkar2002}.  Consider the Lorentzian spacetime string 
effective action for the orbifold, for concreteness, first to leading order in $\apm$
\begin{eqnarray}%
I = \frac{1}{16\pi G} \int_M \sqrt{-g}\, e^{-2\phi}\, [R +4 (\nabla \phi)^2
-\delta^2(x)V(T)]
+\frac{1}{8\pi G} \int_{\partial{M}} \sqrt{-g}\,e^{-2\phi}\, K\, \label{action}
\end{eqnarray}%
where $K$ is the extrinsic curvature and $\delta^2 (x) V(T)$ denotes a possible
tachyon potential localized at the tip of the cone. A nontrivial point here is that the twisted sector of string theory automatically furnishes tachyons localized at the tip of the cone, and it is natural to consider their dynamics localized at the tip of the cone. By contrast, in canonical gravity there is no analogous reason why the classical Einstein equations would be satisfied for any value of the opening angle of the cone. From the worldsheet perspective, variation of this action generates the sigma model beta functional equations. The beta functional vanishes at the critical points of this action where the classical equations of motion are satisfied resulting in a worldsheet conformal field theory. 

The extrinsic
curvature term in the action above is as usual necessary to ensure that the effective
action reproduces the string equations of motion for variations
$\delta\phi$ and $\delta g$ that vanish at the boundary.
The details of the tachyon potential  are not important except that it supplies a Euclidean 7-brane source term for gravity so that the  Einstein equations are satisfied even with a curvature singularity  $R=\delta^2(x)V(T)$ 
 at $x=0$. The main point is that the above action correctly captures the fact that  the dilaton
equations are satisfied with a constant dilaton, and the bulk
contribution to the action is zero. 

To compute the boundary contribution, note that the boundary has
topology $\bR^8\times {\bf S}^1$. For a cone, the circle ${\bf S}^1$ has
radius $r$ but the angular variable will go from $0$ to $\frac{2\pi}{N}$.  
The extrinsic curvature for the circle equals $1/r$ and thus the
contribution to the tree level action from the boundary term is proportional to 
$\frac{2\pi}{N}$.  The total contribution can be written as
\be\label{tree}
\log (\hat Z^{(0)}(N)) = \frac{A}{ 4 G} (\frac{1}{N} -1)\, .
\ee  
where we have fixed an additive constant to  $-1$ by requiring that the action is zero for $N=1$. We have to also remember a factor of $-i$ in Euclidean continuation of
$\sqrt{-g}$. This formula can now be analytically continued  for complex $N$. Substituting in \eqref{entropy} we conclude that the tree level contribution to the entropy is $A/4G$ equal to the Bekenstein-Hawking entropy. 

Note that  one would not be able to make a similar argument in canonical gravity because a conical space with any opening angle does not solve the equations of motion of canonical gravity and is an  off-shell configuration that does not correspond to a saddle point of the action.

In  conformal field theory, we should worry about  higher order $\apm$
corrections to the effective action. These corrections are dependent on field
redefinitions or equivalently on the renormalization scheme of the
world-sheet sigma model. However, the total contribution of these
corrections to bulk action must nevertheless vanish for the orbifold
because we know that the equations of motion of the dilaton are
satisfied with a constant dilaton \emph{exactly} to all orders  in $\apm$ which implies no source terms for
the dilaton in the bulk. 
Possible higher derivative $\apm$ corrections  to the Gibbons-Hawking term are subleading by dimensional analysis. Thus, the entire contribution to the action
comes from the leading Gibbons-Hawking boundary term above even when the $\apm$ corrections are
taken into account \cite{Dabholkar2002}. This  can be calculated reliably  in a scheme
independent way by using the conical geometry of the exact solution at
the boundary to obtain the tree level expression \eqref{tree}. It would be interesting to investigate if  this boundary term can be captured by a worldsheet computation in the conformal field theory \cite{Kraus2002} or in string field theory \cite{DeLacroix2017, Erbin2021}.

\section{Holography and Finiteness of Entanglement Entropy \label{sec:Finite}}

 Rindler spacetime arises as the near horizon geometry of any black hole with a bifurcate horizon, in particular,  the  eternal  Schwarzschild black hole in anti de Sitter spacetime. The left region $U_L$ and the right region $U_R$ of the two-sided black hole geometry correspond to the left Rindler wedge and the right Rindler wedge respectively. In the limit of a very large $AdS$ and black hole radius, this approximation is arbitrarily good.  The Hartle-Hawking  vacuum $|\Omega\rangle$ of the black hole corresponds to the Minkowski vacuum for the near horizon physics. 
 
 Tracing over the states localized in the left region of the black hole geometry gives rise to the Rindler density matrix for the right region:
 \begin{equation}\label{trace}
 	\rho_R	= \TrH[L] |\Omega\rangle \langle \O |
 \end{equation}
 which is to be identified with the thermal density matrix considered earlier \eqref{Rindlerrho}. The associated von Neumann entropy is the entanglement entropy of our interest.
 
 In the field theory limit for the bulk modes, this entanglement entropy is proportional to the area of the entanglement surface and has the well-known ultra-violet divergence. This divergence is a reflection of the strong correlations at short distances in a local quantum field theory, for example through the derivatives of the field across the entangling surface. In taking the trace \eqref{trace} over the states in the left region, we have implicitly assumed that the total Hilbert space $\cH$ is a direct product $\cH_L \otimes  \cH_R$. This assumption is not quite correct since the modes across the entangling surface on the two sides are strongly correlated. Thus, one cannot really associate factorized Hilbert spaces $\cH_L$ and $\cH_R$ with the regions $U_L$ and $U_R$.

 In a local relativistic quantum field theory,  there is no factorization of Hilbert spaces but there is a factorization of algebras of local observables. One can therefore frame the discussion in the language of local observables within algebraic quantum field theory using the Tomita-Takesaki modular theory \cite{Takesaki1970,Borchers2000}.  With the regions $U_L$ and $U_R$ one can associate von Neumann algebras of local observables $\cA_L$ and $\cA_R$ respectively, rather than the factorized Hilbert spaces $\cH_L $ and $\cH_R$. One can then proceed to define the Rindler Hamiltonian $H_R$ as the modular Hamiltonian associated with the state `cyclic and separating' state  $|\Omega\rangle$ and the algebra $\cA_L$. The entanglement entropy is then defined in terms of the modular Hamiltonian without using factorized Hilbert spaces. 
 
By causality,  the two algebras of operators localized in the causally separated regions $U_L$ and $U_R$ have to be mutually commuting:
 \begin{equation}
 	[ \cA_L, \cA_R ] = 0 \, .
 \end{equation}  
 Therefore, by Shur’s lemma,  $\cA_R$ cannot furnish an irreducible representation on a Hilbert space unless $\cH$ is factorized.  This implies that the algebra of observables restricted to the right rindler wedge is a von Neumann algebra of Type III rather than Type I as would be the case for a purely factorized quantum mechanical system. As  emphasized in \cite{Witten2018}, the divergence in the entanglement entropy in a relativistic local quantum field theory is not a property of the state $|\Omega \rangle$ but rather of the von Neumann algebra  $\cA_R$ which is of Type III and not Type I, in that the entanglement entropy in any typical state will manifest this universal UV divergence.

In quantum gravity, there is no natural notion of algebra of local observables and it is not clear how to pose these questions.  However, within a holographic context, one has the following expectation. 
The holographic boundary dual of this bulk black hole  geometry is the product of  conformal quantum field theories with Hilbert spaces $\overline\cH_L$ and $\overline\cH_R$ of states living on the left and the right boundaries respectively. 
 The full boundary Hilbert space $\overline\cH$ is thus the direct product  $\overline \cH_L \otimes  \overline \cH_R$.  The black hole Hartle-Hawking vacuum $|\Omega\rangle$ which appears like a thermal state to the observers in the right region is represented as a  maximally entangled  thermofield double state $|\overline \Omega\rangle$ in this product Hilbert space \cite{Maldacena2001a,VanRaamsdonk2010,Maldacena2013}. Tracing  over the left boundary states gives the thermal density matrix $\overline\rho_R$ for the CFT on the  right boundary consistent with  the fact that the black hole in the bulk corresponds to a thermal state in the right boundary CFT. 
 
 One can consider the algebra of observables $\overline \cA_R$ of operators acting only on $\overline\cH_R$. 
 Since the boundary Hilbert space $\overline\cH$ is a direct product, the algebra of boundary observables $\overline \cA_R$ acting on $\overline\cH_R$ is  a von Neumann algebra of Type I much like for a quantum mechanical system and furnishes an irreducible representation on $\overline H_R$. One can define the modular Hamiltonian $\overline H_R$ corresponding to this algebra and the cyclic separating state $|\overline \Omega\rangle $. The modular density matrix $\overline\rho_R$ is then defined by
\begin{equation}
	\overline \rho_R := e^{-2\pi \overline H_R} \,  
\end{equation}
which coincides with the density matrix \eqref{right-density}.
The entanglement entropy $\overline S$ of our interest is the von Neumann entropy of this density matrix. As argued in \cite{Jafferis2016,Leutheusser2021a,Witten2021a},  $\overline S$ of the boundary theory should be identified with the \textit{total} bulk  entropy $S$ including the tree level contribution from the classical Bekenstein-Hawking entropy.

There is no divergence in the von Neumann entropy for the density matrix $\overline\rho_R$ which is simply the thermal density matrix for one copy of $\cN=4$ super Yang-Mills theory.
 How can this be consistent with the bulk?   In particular,  $\overline\rho_R$ cannot possibly be identified with $\rho_R$ defined in \eqref{trace} which has divergent von Neumann entropy.  As noted earlier, in the field theory limit in the bulk, there is no well-defined factorization $\cH_L \otimes \cH_R$ and the divergence in the entanglement entropy in the field theory limit is a reflection of this lack of factorization. In particular, the  algebra of observables on  $\cA_R$ is a von Neumann algebra of Type III and cannot be identified with the algebra of observables $\overline\cA_R$. 
 
  Holography suggests that quantum gravity must improve the situation. The divergence in the quantum field theory limit of the bulk should be ameliorated in the full quantum string theory in the bulk  so that the behavior in the bulk  correctly corresponds to the behavior in the boundary expected for a Type-I von Neumann algebra of observables, even thought there is no natural notion of local observables in quantum gravity. In particular, 
  \begin{equation}
  	\overline S = S = \frac{A}{4G} + S_q \, .
  \end{equation}
  There is also a corresponding relation between the modular operators:
  \begin{equation}
  	\overline H_R = \frac{\hat A}{4G} + H_R
  \end{equation}
  where $\hat A$ is the horizon area operator \cite{Jafferis2016}.
  
Since the boundary entanglement entropy $\overline S$ is finite, the bulk entanglement entropy $S$ should also be finite order by order in string perturbation theory corresponding to the large $N$ expansion in the boundary. In  Type-II superstring theory with $32$ supercharges, there is no renormalization of Newton's constant to one-loop order and consequently there are no UV or IR divergences in $S^{(0)}$. One therefore expects that the one-loop entropy $S^{(1)}$ \eqref{entropy} in this theory would also not have any divergences. 
 
Recent investigations in \cite{Leutheusser2021,Witten2021b,Leutheusser2021a,Witten2021a} support this reasoning. Instead of asking the harder question of how the bulk quantum string theory improves the divergences of quantum field theory, one could ask how the boundary mimics the divergence of the quantum fields in the bulk. It was shown in \cite{Leutheusser2021,Leutheusser2021a} that a Type-III von Neumann algebra of observables emerges from the Type-I algebra in the boundary theory at large $N$ which can naturally correspond to the algebra $\cA_R$ of local fields  in the bulk. This algebra was identified  as the `small algebra' of generalized free fields in \cite{Papadodimas2012,Papadodimas2013,Papadodimas2013c}. For related work on the reconstruction of bulk observables and the black hole inside see \cite{Hamilton2005,Hamilton2006, Jafferis2020, Almheiri2015, Harlow2018}. The Type-III character of this algebra is related to the ultra-violet divergence in the entropy in the field theory limit. It was shown in \cite{Witten2021a} that corrections subleading in $N$ modify the character of this algebra rendering it Type-II to `improve' the situation. 

These results indicate that the divergence of the entanglement entropy in the bulk is an artifact of the large $N$ limit and should be ameliorated once string loop corrections are taken into account. 
For example, in the context of $AdS_5/CFT_4$, the tree-level Bekenstein entropy of Schwarzschild black hole in the bulk \cite{Gubser1996} would correspond with the  entropy of thermal gas in the super Yang Mills theory on the boundary at large $N$ at large 't Hooft parameter. In the boundary theory, the correction to this entropy at subleading order in $N^2$ is expected to be finite and nonzero and in principle computable. 
Large 't Hooft parameter is the  limit of large $AdS$ radius.  In the large $AdS$ radius limit, $AdS$ spacetime reduces to flat Minkowski spacetime. Therefore, one expects a finite and nonzero value for the quantity  $S_q$ \eqref{q-entropy} of our interest in flat space after appropriate renormalizations.  

Given these arguments,  one is  encouraged to seek a  summation of the series \eqref{entropy-density} for the entropy density and  a finite answer for the entanglement entropy \eqref{entropy} to one-loop order. Modular invariance is a very stringent constraint in string theory. It is thus nontrivial that we are able to obtain an expression for the entropy density that is  manifestly modular invariant. As we have seen, the orbifold method naturally allows a definition to all orders in perturbation theory including the tree level contribution. The generalized entropy thus defined is the natural object to which the generalized second law of thermodynamics can apply \cite{Wall2009, Wall2012}. 

It remains to be seen if the resulting entropy is finite even after the integral over the moduli space. 
While the considerations above rely on properties of algebras of local observables in the boundary CFT, their formulation is expected to involve  very different physical concepts in the bulk string theory, considering that there are   no strictly local observables in  quantum gravity. 

\section*{Acknowledgements}

I would like to thank Kyriakos Papadodimas, Upamanyu Moitra, Ashoke Sen, Edward Witten, and Don Zagier for useful discussions. 

\bibliographystyle{JHEP}
\bibliography{library}

\end{document}